\documentclass[aps,prl,twocolumn,showpacs,preprintnumbers,amsmath,amssymb,floatfix]{revtex4}
\usepackage{graphicx}
\usepackage{dcolumn}
\usepackage{bm}

\newcommand {\beq} {\begin{equation}}
\newcommand {\eeq} {\end{equation}}
\newcommand {\beqa}{\begin{eqnarray}}
\newcommand {\eeqa}{\end{eqnarray}}
\newcommand {\n} {\nonumber}

\newcommand {\tr}{{\rm tr\,}}

\begin{document}


\title{Expanding (3+1)-dimensional universe from a Lorentzian matrix model\\
for superstring theory in (9+1)-dimensions}

\author{Sang-Woo Kim$^{1}$}
\email{sang@het.phys.sci.osaka-u.ac.jp}
\author{Jun Nishimura$^{2,3}$}
\email{jnishi@post.kek.jp}
\author{Asato Tsuchiya$^{4}$}
\email{satsuch@ipc.shizuoka.ac.jp}

\affiliation{
$^{1}$Department of Physics, Osaka University,
Toyonaka, Osaka 560-0043, Japan\\
$^{2}$KEK Theory Center, High Energy Accelerator Research Organization,
		Tsukuba 305-0801, Japan \\
$^{3}$Department of Particle and Nuclear Physics,
School of High Energy Accelerator Science,
Graduate University for Advanced Studies (SOKENDAI),
Tsukuba 305-0801, Japan \\
$^{4}$Department of Physics, Shizuoka University,
836 Ohya, Suruga-ku, Shizuoka 422-8529, Japan
}

\date{August 2011; preprint: KEK-TH-1484, OU-HET-720-2011}


\begin{abstract}
We reconsider the matrix model formulation of
type IIB superstring theory in (9+1)-dimensional space-time.
Unlike the previous
works
in which
the Wick rotation was used to make the model well defined,
we regularize the Lorentzian model
by introducing infrared cutoffs
in both the spatial and temporal directions.
Monte Carlo studies reveal that
the two cutoffs can be removed in the large-$N$ limit
and that the theory thus obtained
has no parameters other than one scale parameter.
Moreover, we find that three out of nine spatial directions
start to
expand at some ``critical time'', after which
the space has SO(3) symmetry instead of SO(9).
\end{abstract}


\maketitle
\paragraph*{Introduction.---}

One of the most fundamental questions concerning our universe is
why we live in a (3+1)-dimensional space-time, and
why the universe is expanding.
The aim of this Letter 
is to provide some evidence that
these facts can be 
derived from
a nonperturbative formulation 
of superstring theory in (9+1) dimensions based on matrix models.
%
%
%
Motivated by recent 
developments in understanding the dynamics
of the 
Euclideanized model,
we study
the ${\rm SO}(9,1)$ symmetric
Lorentzian model 
nonperturbatively
without Wick rotation.
Our Monte Carlo results demonstrate, among other things, 
that three out of nine spatial directions
start to expand
in the early universe.
We expect that
what we are doing here is
essentially a first-principle calculation of
the unified theory
including quantum gravity.
This may be contrasted with
the quantum cosmology in the early 1980s that
aimed at describing the
birth of the universe \cite{vilenkin}
within the mini-superspace approximation.
More recently, a nonperturbative approach to quantum gravity
has been pursued
using the causal dynamical triangulation \cite{Ambjorn:2005qt}.
%
%
%
%
For earlier works that put forward the idea to use matrices
for cosmology, see Refs.~\cite{Freedman:2004xg,mst}.

\paragraph*{Matrix model for superstrings.---}


Superstring theory not only provides a most natural candidate for a
consistent theory of quantum gravity but also enables a unified
description of all the interactions and the matters.
A crucial problem is that
we do not yet have a well-established nonperturbative formulation,
which would be needed in addressing dynamical issues
such as the determination of space-time dimensionality \cite{endnote4d}.

%
%
%


In the 1990s, there was remarkable progress in understanding the
nonperturbative aspects of superstring theory based on
D-branes.
Most importantly, it was noticed that large-$N$ matrices
are the appropriate
microscopic degrees of freedom which are useful in formulating
superstring theory in a nonperturbative
manner \cite{BFSS,IKKT,Dijkgraaf:1997vv}.
In particular, the type IIB matrix model
was proposed as a nonperturbative formulation of
type IIB superstring theory
in ten-dimensional space-time \cite{IKKT}.
%
It was also realized that the five types of
superstring theory in ten dimensions
are just different descriptions of the same theory.
Therefore,
it was speculated
that the type IIB matrix model
actually describes the unique underlying theory, although
it takes the form
that has
explicit connection
to perturbative type IIB superstring theory.

In the type IIB matrix model, the space-time is represented 
\emph{dynamically}
by the eigenvalue distribution
of ten bosonic $N\times N$
traceless Hermitian matrices \cite{AIKKT}.
%
So far, the dynamical generation of four-dimensional space-time
has been discussed exclusively in the Euclideanized model.
Indeed
the spontaneous symmetry breaking (SSB) of SO(10) down to SO(4)
was suggested by
the Gaussian expansion method \cite{SSB,Kawai}.
Recently, systematic calculation of the free energy has been
performed for SO($d$) symmetric
vacua with $2 \le d \le 7$,
and it is found that $d=3$ gives the minimum \cite{Nishimura:2011xy}.
Furthermore, the ratio of the space-time extent
in the extended directions to that in the shrunken directions
is shown to be finite.
%
These results, if true,
suggest the necessity for reconsidering the formulation
in order to make any connection to the real world.
%

\paragraph*{Matrix model with SO(9,1) symmetry.---}

Our starting point is the action $S = S_{\rm b} + S_{\rm f}$, where \cite{IKKT}
\beqa
S_{\rm b} &=& -\frac{1}{4g^2} \, \tr \Bigl( [A_{\mu},A_{\nu}]
[A^{\mu},A^{\nu}] \Bigr) \ , \n  \\
S_{\rm f}  &=& - \frac{1}{2g^2} \,
\tr \Bigl( \Psi _\alpha (\, {\cal C} \,  \Gamma^{\mu})_{\alpha\beta}
[A_{\mu},\Psi _\beta] \Bigr)  \ ,
\label{action}
\eeqa
with  $A_\mu$ ($\mu = 0,\cdots, 9$) and
$\Psi_\alpha$ ($\alpha = 1,\cdots , 16$) being
$N \times N$ traceless Hermitian matrices.
The Lorentz indices $\mu$ and $\nu$ are contracted using the metric
$\eta={\rm diag}(-1 , 1 , \cdots , 1)$.
The $16 \times 16$ matrices $\Gamma ^\mu$ are
ten-dimensional gamma matrices after the Weyl projection,
and the unitary matrix ${\cal C}$ is the
charge conjugation matrix.
The action has manifest SO(9,1) symmetry,
where $A_{\mu}$ and $\Psi _\alpha$ transform as a
vector and
a Majorana-Weyl spinor, respectively.
The Euclidean model,
which has SO(10) symmetry,
can be obtained from this action
by the Wick rotation $A_0=i A_{10}$.
A crucial difference is that
the bosonic part of the action in the Euclidean model
is positive definite, whereas in the Lorentzian model it is
\beq
\tr (F_{\mu\nu}F^{\mu\nu})=
- 2 \, \tr (F_{0i})^2 + \tr (F_{ij})^2 \ ,
\label{bosonic-action}
\eeq
where $F_{\mu\nu} = - i [A_\mu , A_\nu]$ are Hermitian matrices,
and hence the two terms in (\ref{bosonic-action}) have opposite 
signs \cite{endnote2}.

We study, for the first time, the Lorentzian model nonperturbatively
based on the partition function
\beq
Z = \int d A \, d\Psi \, e^{i S} =
\int d A \,  e^{i S_{\rm b}} {\rm Pf}{\cal M}(A)
\ ,
\label{partition-fn-def}
\eeq
where the Pfaffian ${\rm Pf}{\cal M}(A)$ appears from integrating
out the fermionic matrices $\Psi_\alpha$.
Note that in the Euclidean model,
the Pfaffian is complex in general,
and its phase plays a crucial role in the aforementioned SSB
of SO(10) symmetry \cite{NV,Anagnostopoulos:2001yb}.
On the other hand, the Pfaffian in the Lorentzian model is \emph{real}.
Therefore, the mechanism of SSB that was identified
in the Euclidean model is absent in the Lorentzian model.

In the definition (\ref{partition-fn-def}),
we have replaced the ``Boltzmann weight'' $e^{-S}$
used in the Euclidean model by $e^{iS}$.
This is theoretically motivated
from the connection
to the worldsheet theory \cite{IKKT}.
The partition function
(\ref{partition-fn-def})
can also be obtained formally
from pure ${\cal N}=1$ supersymmetric Yang-Mills theory
in $(9+1)$ dimensions by dimensional reduction.
Note, however, that
the expression (\ref{partition-fn-def}) is ill defined and
requires appropriate regularization
in order
to make any sense out of it.
This is in striking contrast to the Euclidean model,
in which the partition function is shown to
be finite without any regularization \cite{Krauth:1998xh,AW}.

It turns out that the integration over $A_0$ is divergent,
even if we fix $\frac{1}{N} \tr (A_i)^2$ to a constant.
In order to cure this divergence,
we introduce a constraint
\beq
\frac{1}{N}\tr (A_0)^2  \le  \kappa \frac{1}{N} \tr (A_i)^2  \ ,
\label{T-constr}
\eeq
which is invariant under the scale transformation
$A_\mu \rightarrow \rho A_\mu$.
Note that this constraint generically breaks SO(9,1) symmetry down to SO(9).
However, it turns out to be equivalent to 
imposing (\ref{T-constr}) after
``gauge fixing'' the boost symmetry by requiring that
$\frac{1}{N}\tr (\tilde{A}_0)^2$ with
$\tilde{A}_\mu = O_{\mu \nu} A_\nu$
be minimized with respect to $O \in {\rm SO}(9,1)$.
In this sense, the constraint actually
respects the SO(9,1) symmetry.


Let us note
that $e^{iS_{\rm b}}$ in the partition function (\ref{partition-fn-def})
is a phase factor
just as in the path-integral formulation of quantum field theories
in Minkowski space.
As is commonly done in integrating oscillating functions,
we introduce the convergence factor
$e^{- \epsilon |S_{\rm b}|}$ and take the $\epsilon \rightarrow 0$ limit
after the integration.
%

The partition function can then be rewritten as
\begin{align}
Z &= \int dA \int_0^{\infty} dr \,
\delta\left(\frac{1}{N}\tr (A_i)^2 -r \right)
 e^{i S_{\rm b}-\epsilon |S_{\rm b}| }
{\rm Pf} {\cal M}  \ , \nonumber
%
\end{align}
where the integration over $A_\mu$ is assumed to be
restricted by
the constraint (\ref{T-constr}).
%
%
%
After rescaling the variables $A_\mu \rightarrow r^{1/2} A_\mu$
in the integrand,
we integrate over $r$ and get
\beq
\int_0^\infty  dr  \, r^{\frac{18}{2}(N^2-1)-1}
 e^{r^2 (i S_{\rm b}-\epsilon |S_{\rm b}|) }
\propto \frac{1}{|S_{\rm b}|^{\frac{18}{4}(N^2-1)}} \ ,
\label{integrate-r}
\eeq
which diverges for $S_{\rm b}=0$.
In order to cure this divergence, we introduce a constraint
\beq
\frac{1}{N} \tr (A_i)^2   \le  L^2
\label{R-constr}
\eeq
before the rescaling.
Then the integration domain for $r$ becomes $[0,L^2]$, and 
(\ref{integrate-r}) is replaced by
$f(S_{\rm b})$, where $f(x)$ is a function with a sharp peak at $x=0$.
Thus we arrive at the model
\begin{align}
Z = & \int dA \,
f\left(
\frac{1}{N}\tr (F_{\mu\nu}F^{\mu\nu}) - C \right)
 {\rm Pf} {\cal M} (A)
 \nonumber \\
& \times
\delta\left(\frac{1}{N}\tr (A_i)^2 - 1 \right)
\theta\left(\kappa - \frac{1}{N}\tr (A_0)^2  \right)  \ ,
\label{our-model}
\end{align}
where $\theta(x)$ is the Heaviside step function.
The constant $C$ should be
set to zero according to
our derivation.
If we consider the $C<0$ case, the model (\ref{our-model})
may be viewed as the matrix model motivated from
the space-time uncertainty principle \cite{Yoneya:1997gs}
with the regularization in the second line,
which we find to be necessary.
Since the Pfaffian ${\rm Pf} {\cal M}(A)$ is real
in the present Lorentzian case,
the model (\ref{our-model}) can be studied by Monte Carlo simulation
without the sign problem.
Note that this is usually not the case for quantum field theories in
Minkowski space.
\paragraph*{Monte Carlo results.---}

We perform Monte Carlo simulation of the model (\ref{our-model})
with $C=0$
by using the Rational Hybrid Monte Carlo algorithm \cite{Clark:2004cp},
which is quite standard in recent simulations of 
quantum chromodynamics including the effects of dynamical quarks.

\begin{figure}[htb]
\begin{center}
\includegraphics[height=6cm]{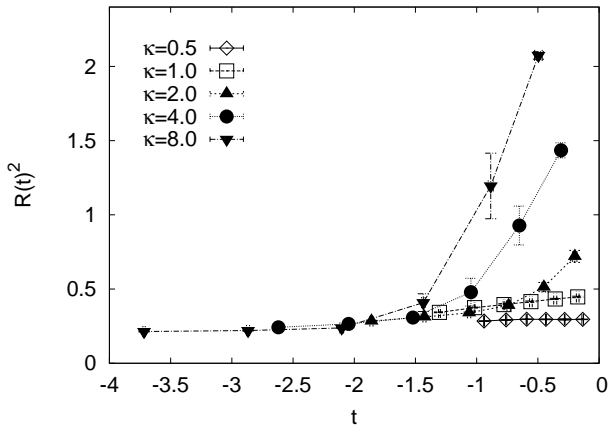}
\end{center}
\caption{
The extent of space $R(t)^2$ with $N=16$ and $n=4$
is plotted as a function of $t$
for five values of $\kappa$.
The peak at $t=0$ starts to grow at some critical $\kappa$.
}
\label{Rt}
\end{figure}

In order to see the time evolution, we diagonalize $A_0$,
and define the eigenvectors $| t_a \rangle$ corresponding
to the eigenvalues $t_a$ of $A_0$ ($a=1 , \cdots , N$)
with the specific order $t_1 < \cdots < t_N$.
The spatial matrix in this basis $\langle t_{a} | A_i | t_{b} \rangle $
is not diagonal, but it turns out that the off-diagonal elements
decrease rapidly as one goes away from a diagonal element.
This motivates us to
define $n\times n$ matrices
$\bar{A}_i^{(ab)}(t) \equiv  \langle t_{\nu+a} | A_i | t_{\nu+b} \rangle $
with $1 \le a , b \le n$ and
$t= \frac{1}{n}\sum_{a=1}^{n} t_{\nu + a}$
for $\nu=0,\cdots , (N-n)$.
These matrices represent the 9d space structure 
at fixed time $t$ \cite{endnote-t}.
The block size $n$ should be large enough to include non-negligible 
off-diagonal elements.
In Fig.~\ref{Rt} we plot the extent of space 
$R(t)^2 \equiv \frac{1}{n} \tr
\bar{A}_i(t)
^2$ for $N=16$ and $n=4$.
Since the result is symmetric under the time reflection
$t \rightarrow -t$ as a consequence of the symmetry $A_0 \rightarrow -A_0$,
we only show the results for $t<0$.
There is a critical $\kappa$, beyond which
the peak at $t=0$ starts to grow.

\begin{figure}[htb]
\begin{center}
\includegraphics[height=6cm]{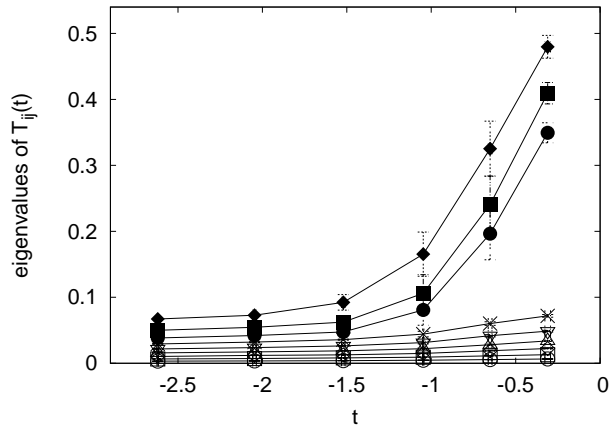}
\end{center}
\caption{
The nine eigenvalues of $T_{ij}(t)$ with $N=16$ and $n=4$
are plotted as a function of $t$
for $\kappa=4.0$. After the critical time $t_{\rm c}$,
three eigenvalues become larger,
suggesting that the SO(9) symmetry is spontaneously broken
down to SO(3).
}
\label{Tij-t}
\end{figure}

Next we study the spontaneous breaking of the SO(9) symmetry.
As an order parameter,
we define the $9 \times 9$
(positive definite)
real symmetric tensor
\beq
T_{ij}(t) = \frac{1}{n} \tr \Bigl\{
\bar{A}_i(t) \bar{A}_j(t) \Bigr\} \ ,
\eeq
which is an analog of the moment of inertia tensor.
The nine eigenvalues of $T_{ij}(t)$ are
plotted against $t$ in Fig.~\ref{Tij-t} for $\kappa=4.0$.
We find that three largest eigenvalues of
$T_{ij}(t)$ start to grow at the critical
time $t_{\rm c}$, which suggests that the SO(9)
symmetry is spontaneously broken down to
SO(3) after $t_{\rm c}$.
Note that $R(t)^2$ is given by the sum of nine eigenvalues
of $T_{ij}(t)$.

\paragraph*{Mechanism of the SSB.---}

The SSB of SO(9) looks mysterious at first sight, but
we can actually understand the mechanism quite intuitively.
Let us consider the case in which $\kappa$ is large.
Then the first term of (\ref{bosonic-action}) becomes
a large negative value, and therefore the second term
has to become large in order to 
make (\ref{bosonic-action}) close to zero as required in (\ref{our-model}).
Because of
the constraint $\frac{1}{N} \tr (A_i)^2=1$, however,
it is more efficient to maximize the second term of (\ref{bosonic-action})
at some fixed time.
The system actually chooses the middle point $t=0$, where
the suppression on $A_i$ from the first term of (\ref{bosonic-action})
becomes the least.
This explains why the peak of $R(t)$ at $t=0$ grows as we increase $\kappa$.

Let us then consider a simplified question: what is the configuration of $A_i$
which gives the maximum $\frac{1}{N} \tr (F_{ij})^2$ with fixed
$\frac{1}{N} \tr (A_i)^2=1$. Using the Lagrange multiplier $\lambda$,
we maximize the function $G=  \tr (F_{ij})^2 - \lambda \,  \tr (A_i)^2$.
Taking the derivative with respect to $A_i$, we obtain
$2\, [A_j,[A_j, A_i]] - \lambda A_i = 0 $.
This equation can be solved if $A_i = \chi L_i$ for $i\le d$,
and $A_i = 0$ for $d < i \le 9$,
where $L_i$ are the representation matrices
of a compact semi-simple Lie algebra with $d$ generators.
Clearly $d$ should be less than or equal to 9.
It turns out that the maximum of
$\frac{1}{N} \tr (F_{ij})^2$ is achieved for the SU(2) algebra,
which has $d=3$,
with $L_i$ being the direct sum of the spin-$\frac{1}{2}$ representation
and $(N-2)$ copies of the trivial representation.
This implies the SSB of SO(9) down to SO(3).
The SSB
can thus be understood
as a classical effect in the $\kappa\rightarrow \infty$ limit.
%
When we tune $\kappa$ with increasing $N$ as described below,
quantum effects become important.
We have confirmed \cite{KNT} that the $n\times n$ matrix
$Q = \sum_{i=1}^9 \bar{A}_i(t)^2$ has 
quite a continuous eigenvalue distribution,
which implies that the space is not like a two-dimensional sphere as
one might
suspect from the classical picture.


\paragraph*{Removing the cutoffs.---} 

It turned out that one can remove the
infrared cutoffs $\kappa$ and $L$ 
in the large-$N$ limit
in such a way that $R(t)$ scales.
This can be done in two steps.
(i) First we send $\kappa$ to $\infty$ with $N$ as
$\kappa = \beta \, N^{p}$ ($p\simeq \frac{1}{4}$) \cite{KNT}.
The scaling curve of $R(t)$ one obtains in this way
depends on $\beta$.
(ii) Next we send $\beta$ to $\infty$ with $L$.
The two limits correspond to
the continuum limit and
the infinite volume limit, respectively, in quantum field theory.
Thus the two constraints (\ref{T-constr}), (\ref{R-constr})
can be removed in the large-$N$ limit,
and 
the resulting theory has no parameter other than one scale parameter.

\begin{figure}[htb]
\begin{center}
\includegraphics[height=6cm]{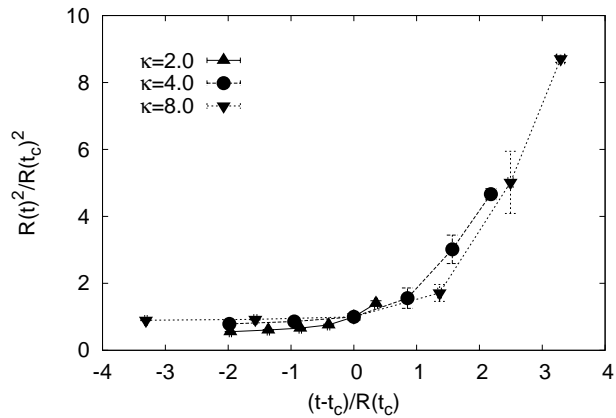}
\end{center}
\caption{
Our data for $R(t)^2$ shown in Fig.~\ref{Rt} 
with $\kappa$ larger than the critical value
are replotted against the
shifted time $t - t_{\rm c}$ in units of
the size of the universe $R(t_{\rm c})$
at the critical time.
}
\label{Rt-rescaled}
\end{figure}

Let us discuss the second limit (ii) in more detail.
We find that the inequality (\ref{R-constr}) is actually saturated 
for the dominant configurations.
Therefore, one only has to make the rescaling $A_\mu \mapsto L A_\mu$
in order to translate the configurations in the model (\ref{our-model})
as those in the original partition function.
It turns out that $R(t)$ for the rescaled configurations scales
in $\beta$ by tuning $L$ and shifting $t$ appropriately.
In order to see this, it is convenient to 
choose $L$ so that $R(t)$ 
at the critical time $t=t_{\rm c}$ becomes unity,
and to shift $t$ so that the critical time comes to the origin.
Then $R(t)$ with increasing $\beta$ 
extends in $t$
in such a way that the results at smaller $|t|$ scale.
This is demonstrated in Fig.~\ref{Rt-rescaled},
where we find a reasonable scaling behavior
for $N=16$ with $\kappa=2.0, 4.0, 8.0$.
In fact, supersymmetry of the model plays an important
role here \cite{KNT}.

\paragraph*{Summary.---} 
In this Letter
we have studied the nonperturbative dynamics
of the Lorentzian matrix model for 
type IIB superstring theory in ten dimensions.
In order to 
make the model well defined,
we
introduce
the infrared cutoffs on both the spatial and temporal directions.
We find that the two cutoffs can be removed in the large-$N$ limit.
Moreover, the theory thus obtained has no parameters other than one
scale parameter, 
which is a property expected for nonperturbative
superstring theory.
The SO(9) symmetry breaks down to SO(3)
at some critical time, and
the size of the three-dimensional space increases with time.
The cosmological singularity is 
naturally avoided 
due to noncommutativity.
%
%
%

There
are a lot of questions that should be addressed
in our model.
One of the most urgent questions is
whether a
local field theory on
a commutative space-time appears
at the low energy scale.
A possible way is to calculate correlation functions
of the Wilson loop operators \cite{Fukuma:1997en}.
If this question is answered in the affirmative, we consider it very likely
that our model really describes the birth of our universe
from first principles.
The next step would be 
to show
that the four fundamental interactions and the matter fields
appear in our universe at a later time.



\paragraph*{Acknowledgments.---}
We thank H.~Aoki, S.~Iso,
H.~Kawai, Y.~Kitazawa, Y.~Sekino and S.~Yamaguchi
for discussions.
Computations were carried out on
SR16000 at Yukawa Institute.
S.-W.K.\ is supported by Grant-in-Aid
for Scientific Research
from the Ministry of
Education, Culture, Sports, Science and Technology in Japan (No. 20105002).
J.N.\ and A.T.\ are supported in part by Grant-in-Aid
for Scientific Research
(No.\ 19340066, 19540294, 20540286 and 23244057)
from JSPS.





\begin{thebibliography}{99}
\bibitem{endnote-t}
This point of view can be justified in the large-$N$ limit,
in which more and more eigenvalues $\alpha$
of $A_0$ appear within a fixed range $ |\alpha-t| \le \delta t $.
\end{thebibliography}
\end{document}